\newcommand{\xba}{\alpha}

\newcommand{\xbe}{\in}
\newcommand{\xbf}{\phi}

\newcommand{\xbm}{\mu}

\newcommand{\xCN}{\neg}

\newcommand{\xCQ}{\emptyset}

\newcommand{\xcE}{\exists}

\newcommand{\xcc}{\subseteq}

\newcommand{\xcj}{\Leftrightarrow}

\newcommand{\xcm}{\models}

\newcommand{\xcp}{\rightarrow}

\newcommand{\xcs}{\cap}

\newcommand{\xDH}{\item }

\newcommand{\xda}{{\cal A}}

\newcommand{\xdf}{{\cal F}}

\newcommand{\xdl}{{\cal L}}

\newcommand{\xdp}{{\cal P}}

\newcommand{\xEI}{\begin{itemize}}
\newcommand{\xEJ}{\end{itemize}}

\newcommand{\xEd}{\neq}

\newcommand{\xEh}{\begin{enumerate}}

\newcommand{\xEj}{\end{enumerate}}

\newcommand{\xEn}{\begin{description}}
\newcommand{\xEp}{\end{description}}

\newcommand{\xeb}{\prec}

\newcommand{\Xl}{\ldots}

\newcommand{\bl}{\begin{lemma} \rm}
\newcommand{\el}{\end{lemma}}
\newcommand{\br}{\begin{remark} \rm}
\newcommand{\er}{\end{remark}}
\newcommand{\be}{\begin{example} \rm}
\newcommand{\ee}{\end{example}}
\newcommand{\bco}{\begin{corollary} \rm}
\newcommand{\eco}{\end{corollary}}
\newcommand{\bc}{\begin{claim} \rm}
\newcommand{\ec}{\end{claim}}
\newcommand{\bfa}{\begin{fact} \rm}
\newcommand{\efa}{\end{fact}}
\newcommand{\bp}{\begin{proposition} \rm}
\newcommand{\ep}{\end{proposition}}
\newcommand{\bd}{\begin{definition} \rm}
\newcommand{\ed}{\end{definition}}
\newcommand{\bcs}{\begin{construction} \rm}
\newcommand{\ecs}{\end{construction}}
\newcommand{\bcd}{\begin{condition} \rm}
\newcommand{\ecd}{\end{condition}}
\newcommand{\bt}{\begin{theorem} \rm}
\newcommand{\et}{\end{theorem}}
\newcommand{\bn}{\begin{notation} \rm}
\newcommand{\en}{\end{notation}}
\newcommand{\bfi}{\begin{bild} \rm}
\newcommand{\efi}{\end{bild}}
\newcommand{\bsta}{\begin{statement} \rm}
\newcommand{\esta}{\end{statement}}
\newcommand{\bcom}{\begin{comment} \rm}
\newcommand{\ecom}{\end{comment}}
\newcommand{\bdia}{\begin{diagram} \rm}
\newcommand{\edia}{\end{diagram}}

\newcommand{\bfc}{\begin{figure}[htb] \begin{center}}
\newcommand{\efc}{\end{center} \end{figure}}

\documentclass{article}

\usepackage{amssymb,latexsym,epic,eepic,rotating}

\title{A Short Remark on Analogical Reasoning
\thanks{File: ana
}
}

\author{Karl Schlechta
\thanks{
schcsg@gmail.com - https://sites.google.com/site/schlechtakarl/ -
Koppeweg 24, D-97833 Frammersbach, Germany}
\thanks{
Retired, formerly: Aix-Marseille Universit\'{e}, CNRS, LIF UMR 7279, F-13000
Marseille, France
}
}

\begin{document}

\newtheorem{lemma}{Lemma}[section]
\newtheorem{theorem}[lemma]{Theorem}
\newtheorem{proposition}[lemma]{Proposition}
\newtheorem{corollary}[lemma]{Corollary}
\newtheorem{claim}[lemma]{Claim}
\newtheorem{fact}[lemma]{Fact}
\newtheorem{remark}[lemma]{Remark}
\newtheorem{definition}{Definition}[section]
\newtheorem{construction}{Construction}[section]
\newtheorem{condition}{Condition}[section]
\newtheorem{example}{Example}[section]
\newtheorem{notation}{Notation}[section]
\newtheorem{bild}{Figure}[section]
\newtheorem{comment}{Comment}[section]
\newtheorem{statement}{Statement}[section]
\newtheorem{diagram}{Diagram}[section]

\renewcommand{\labelenumi}
  {(\arabic{enumi})}
\renewcommand{\labelenumii}
  {(\arabic{enumi}.\arabic{enumii})}
\renewcommand{\labelenumiii}
  {(\arabic{enumi}.\arabic{enumii}.\arabic{enumiii})}
\renewcommand{\labelenumiv}
  {(\arabic{enumi}.\arabic{enumii}.\arabic{enumiii}.\arabic{enumiv})}

\maketitle

\setcounter{secnumdepth}{3}
\setcounter{tocdepth}{3}

\begin{abstract}

We discuss the problem of defining a logic for
analogical reasoning, and sketch a solution in the style of the semantics for
Counterfactual Conditionals, Preferential Structures, etc.

\end{abstract}

\tableofcontents
\clearpage

%
%
%


\section{
Introduction
}

We consider here (largely verbatim, only punctually slightly modified)
excerpts
from  \cite{SEP13}, see also  \cite{SEP19c}, to set the stage
for
Section \ref{Section Idea} (page \pageref{Section Idea}).
\subsection{
Section 2.2, p. 5 of \cite{SEP13}
}

\bd

$\hspace{0.01em}$


\label{Definition}

An analogical argument has the following
form:

1. $S$ is similar to $T$ in certain (known) respects.

2. $S$ has some further feature $Q.$

3. Therefore, $T$ also has the feature $Q,$ or some feature $Q*$
similar to $Q.$

(1) and (2) are premises. (3) is the conclusion of the argument.
The argument form is
inductive; the conclusion is not guaranteed to follow from the
premises.

\ed

$S$ and $T$ are referred to as the source domain and target domain,
respectively. A domain is
a set of objects, properties, relations and functions, together
with a set of accepted
statements about those objects, properties, relations and
functions. More formally, a
domain consists of a set of objects and an interpreted set of
statements about them. The
statements need not belong to a first-order language, but to keep
things simple, any
formalizations employed here will be first-order. We use
unstarred symbols (a, $P,$ $R,$ $f)$ to
refer to items in the source domain and starred symbols $(a*,$ $P*,$
$R*,$ $f*)$ to refer to
corresponding items in the target domain.

\bd

$\hspace{0.01em}$


\label{Definition Mapping}

Formally, an analogy between $S$ and $T$ is a one-to-one mapping
between objects,
properties, relations and functions in $S$ and those in $T.$
\subsection{
Section 2.2, pp. 6-7 of \cite{SEP13}
}

\ed

In an earlier discussion of analogy, Keynes,
in  \cite{Key21}, introduced
some terminology that is
also helpful.
 \xEh
 \xDH
Positive analogy.

Let $P$ stand for a list of accepted propositions $P1, \Xl,$ Pn about
the source domain $S.$
Suppose that the corresponding propositions $P*1, \Xl,$ $P*n,$
abbreviated as $P*,$ are
all accepted as holding for the target domain $T,$ so that $P$ and $P*$
represent accepted
(or known) similarities. Then we refer to $P$ as the positive
analogy.

 \xDH
Negative analogy.

Let A stand for a list of propositions $A1, \Xl,$ Ar accepted as
holding in $S,$ and $B*$ for
a list $B1*, \Xl,$ $Bs*$ of propositions holding in $T.$ Suppose that the
analogous
propositions $A*$ $=$ $A1*, \Xl,$ $Ar*$ fail to hold in $T,$ and
similarly
the propositions $B$ $=$
$B1, \Xl,$ Bs fail to hold in $S,$ so that A, $ \xCN A*$ and $ \xCN B,$
$B*$ represent
accepted (or
known) differences. Then we refer to A and $B$ as the negative
analogy.

 \xDH
Neutral analogy.

The neutral analogy consists of accepted propositions about $S$ for
which it is not
known whether an analogue holds in $T.$

 \xDH
Hypothetical analogy.

The hypothetical analogy is simply the proposition $Q$ in the
neutral analogy that is
the focus of our attention.

 \xEj

These concepts allow us to provide a characterization for an
individual analogical
argument that is somewhat richer than the original one.

\bd

$\hspace{0.01em}$


\label{Definition Aug}

(Augmented representation)

Correspondence between SOURCE (S) and TARGET (T)

 \xEh

 \xDH
Positive analogy:

$P$ $ \xcj $ $P*$
 \xDH
Negative analogy:

A $ \xcj $ $ \xCN A*$

and

$ \xCN B$ $ \xcj $ $B*$
 \xDH
Plausible inference:

$Q$ $ \xcj $ $Q*$
 \xEj

An analogical argument may thus be summarized:
It is plausible that $Q*$ holds in the target because of certain
known (or
accepted) similarities with the source domain, despite certain
known (or
accepted) differences.
\subsection{
Section 2.4 of \cite{SEP13}
}

\ed

Scepticism:

Of course, it is difficult to show that no successful analogical
inference rule will ever be
proposed. But consider the following candidate, formulated using
the concepts of the schema
in Definition \ref{Definition Aug} (page \pageref{Definition Aug})
and taking us only a short step beyond that basic
characterization.

\bd

$\hspace{0.01em}$


\label{Definition Scept}

Suppose $S$ and $T$ are the source and target domains. Suppose $P1, \Xl,$
Pn (with $n$ $=$ 1)
represents the positive analogy, $A1, \Xl,$ Ar and $ \xCN B1, \Xl,$ $
\xCN Bs$
represent the (possibly
vacuous) negative analogy, and $Q$ represents the hypothetical
analogy. In the
absence of reasons for thinking otherwise, infer that $Q*$ holds in
the target domain
with degree of support $p$ $>$ 0, where $p$ is an increasing function
of $n$ and a
decreasing function of $r$ and $s.$

(Definition \ref{Definition Scept} (page \pageref{Definition Scept})
is modeled on the straight rule for enumerative
induction and inspired by Mill's
view of analogical inference, as described
in  \cite{SEP13} above. We use the
generic phrase ``degree of
support'' in place of probability, since other factors besides the
analogical argument may
influence our probability assignment for $Q*.)$

\ed

It is pretty clear that the schema in
Definition \ref{Definition Scept} (page \pageref{Definition Scept})
is a non-starter. The main problem is
that the rule justifies too
much.

So, how do we chose the ``right one''?
\subsection{
A Side Remark
}

The author was surprised to find a precursor to his concept of
homogenousness in the work of J. M. Keynes,
 \cite{Key21}, quoted in Section 4.3 of  \cite{SEP13}.
\clearpage
\section{
The Idea
}

\label{Section Idea}

We now describe the idea, and compare it to other ideas in philosophical
and AI related logics.

But first, we formalize above ideas into a definition.

\bd

$\hspace{0.01em}$


\label{Definition Ana}

Let $ \xdl $ be an alphabet.
 \xEh
 \xDH
Let $ \xdl_{ \xba } \xcc \xdl,$ and $ \xba: \xdl_{ \xba } \xcp \xdl $ an
injective function,
preserving the type of symbol, e.g.,
 \xEI
 \xDH
if $x \xbe \xdl_{ \xba }$ stands for an object of the universe, then so
will $ \xba (x)$
 \xDH
if $X \xbe \xdl_{ \xba }$ stands for a subset of the universe, then so
will $ \xba (X)$
 \xDH
if $P(.) \xbe \xdl_{ \xba }$ stands for an unary predicate of the
universe,
then so will $ \xba (P)(.)$
 \xDH
etc., also for higher symbols, like $f: \xdp (U) \xcp \xdp (U),$ $U$ the
universe.
 \xEJ
 \xDH
Let $ \xdf_{ \xba }$ a subset of the formulas formed with symbols from $
\xdl_{ \xba }.$

For $ \xbf \xbe \xdf_{ \xba },$ let $ \xba (\xbf)$ be the obvious
formula constructed from $ \xbf $
with the function $ \xba.$
 \xDH
We now look at the truth values of $ \xbf $ and $ \xba (\xbf),$ $v(\xbf
)$ and $v(\xba (\xbf)).$
In particular, there may be $ \xbf $ s.t. $v(\xbf)$ is known, $v(\xba (
\xbf))$ not, and
we extrapolate that $v(\xbf)=v(\xba (\xbf)),$ this is then the
analogical
reasoning based on $ \xba.$

More precisely:
 \xEh
 \xDH
There may be $ \xbf $ s.t. $v(\xbf)$ is not known, $v(\xba (\xbf))$
is known or not, such
$ \xbf $ do not interest us here.

Assume in the following that $v(\xbf)$ is known.
 \xDH
$v(\xbf)$ and $v(\xba (\xbf))$ are known, and $v(\xbf)=v(\xba (
\xbf)).$ The set of such $ \xbf $
is the positive support of $ \xba,$ $ \xba^{+}.$
 \xDH
$v(\xbf)$ and $v(\xba (\xbf))$ are known, and $v(\xbf) \xEd v(\xba
(\xbf)).$ The set of such $ \xbf $
is the negative support of $ \xba,$ $ \xba^{-}.$
 \xDH
$v(\xbf)$ is known, $v(\xba (\xbf))$ is not known. The set of such $
\xbf $
is denoted $ \xba^{?}.$

The ``effect'' of $ \xba $ is to conjecture, by analogy, that $v(\xbf)=v(
\xba (\xbf))$ for
such $ \xbf.$
 \xEj
 \xEj
Intuitively, $ \xba^{+}$ strengthens the case of $ \xba,$ $ \xba^{-}$
weakens it - but these
need not be the only criteria, see also
 \cite{SEP13} and  \cite{SEP19c}.

\ed

Let $ \xda $ be a set of functions $ \xba $ as defined in
Definition \ref{Definition Ana} (page \pageref{Definition Ana}).

We may close $ \xda $ under combinations, as illustrated in the following
Example \ref{Example Combi} (page \pageref{Example Combi}), or not.

\be

$\hspace{0.01em}$


\label{Example Combi}

Consider $ \xba,$ $ \xba'.$

Let $x,x',P,Q$ $ \xbe $ $ \xdl_{ \xba }= \xdl_{ \xba' },$ $Q(x),Q(x')
\xbe \xba^{?}, \xba'^{?}.$
 \xEh
 \xDH
$ \xba $ works well for $x,$ but not for $x':$
$P(x)= \xba (P)(x),$ $P(x') \xEd \xba (P)(x'),$
so $P(x) \xbe \xba^{+},$ $P(x') \xbe \xba^{-},$
 \xDH
$ \xba' $ works well for $x',$ but not for $x:$
$P(x')= \xba' (P)(x'),$ $P(x) \xEd \xba' (P)(x),$
so $P(x) \xbe \xba'^{-},$ $P(x') \xbe \xba'^{+}.$
 \xEj

Let further $ \xba (Q)(x) \xEd \xba' (Q)(x)$ and $ \xba (Q)(x') \xEd
\xba' (Q)(x').$

What shall we do, should we chose one, $ \xba $ or $ \xba',$ for
guessing, or
combine $ \xba $ and $ \xba' $ to $ \xba'',$ chosing $ \xba'' = \xba $
for expressions with $x,$
and $ \xba'' = \xba' $ for expressions with $x',$ more precisely
$ \xba'' (Q)(x):= \xba (Q)(x),$ and $ \xba'' (Q)(x'):= \xba' (Q)(x')$
?

\ee

The idea is now to push the choice of suitable $ \xba \xbe \xda $ into a
relation $ \xeb,$
expressing quality of the analogy.
E.g., in Example 
\ref{Example Combi} (page 
\pageref{Example Combi}), $ \xba'' \xeb \xba $ and $ \xba
'' \xeb \xba' $ - for historical
reasons, smaller elements will be ``better''.

Usually, this ``best'' relation will be partial only, and
there will be many ``best'' $f.$
Thus, it seems natural to conclude the properties which hold
in ALL best $f.$

\bd

$\hspace{0.01em}$


\label{Definition Valid}

Let $ \xda $ be a set of functions as described in
Definition 
\ref{Definition Ana} (page 
\pageref{Definition Ana}), and $ \xeb $ a relation on $ \xda $
(expressing ``better''
analogy wrt. the problem at hand).

We then write $ \xda \xcm_{ \xeb } \xbf $ iff $ \xbf $ holds in all $ \xeb
-$best $f \xbe \xda.$

(This is a sketch only, details have to be filled in according to the
situation considered.)
\subsection{
Discussion
}

\ed

This sounds like cheating: we changed the level of abstraction, and
packed the question of ``good'' analogies into the $ \xeb $-relation.

But when we look at the Stalnaker-Lewis semantics of counterfactual
conditionals, see  \cite{Sta68},  \cite{Lew73},
the preferential semantics for non-monotonic reasoning and
deontic logic, see e.g.  \cite{Han69},  \cite{KLM90},  \cite{Sch04}, 
\cite{Sch18},
the distance semantics for theory revision, see e.g.  \cite{LMS01},
 \cite{Sch04},
this is a
well used ``trick'' we need not be ashamed of.

In above examples, the comparison was between possible worlds, here it is
between usually more complicated structures (functions), but this is no
fundamental difference.

But even if we think that there is a element of cheating in our idea, we
win something: properties which hold in ALL preferential structures,
and which may be stronger for stronger relations $ \xeb,$ see
Fact \ref{Fact Representation} (page \pageref{Fact Representation})  below.
\subsection{
Problems and solutions
}

 \xEh
 \xDH

In the case of infinitely many f's we might have a definability
problem, as the resulting best guess might not be definable any more -
as in the case of preferential structures,
see e.g.  \cite{Sch04}.

 \xDH

Abstract treatment of representation problems for abovementioned logics
work with arbitrary sets, so we have a well studied machinery for
representation results for various types of relations of ``better''
analogies - see e.g.
 \cite{LMS01},  \cite{Sch04},  \cite{Sch18}.

To give the reader an idea of such representation resuls, we mention some,
slightly simplified.

\bd

$\hspace{0.01em}$


\label{Definition Representation}

(1) Let again $ \xeb $ be the relation, and $ \xbm (X):=\{x \xbe X: \xCN
\xcE x' \xbe X.x' \xeb x\},$

(2) $ \xeb $ is called smooth iff for all $x \xbe X,$ either $x \xbe \xbm
(X)$ or there is
$x' \xbe \xbm (X),$ $x' \xeb x,$

(3) $ \xeb $ is called ranked iff for all $x,y,z,$ if neither $x \xeb y$
nor $y \xeb x,$ then
if $z \xeb x,$ then $z \xeb y,$ too, and, analogously, if $x \xeb z,$ then
$y \xeb z,$ too.

\ed

We then have e.g.

\bfa

$\hspace{0.01em}$


\label{Fact Representation}

 \xEh
 \xDH
General and transitive relations are characterised by

$(\xbm \xcc)$ $ \xbm (X) \xcc X$

and

$(\xbm PR)$ $X \xcc Y$ $ \xcp $ $ \xbm (Y) \xcs X \xcc \xbm (X)$
 \xDH
Smooth and transitive smooth relations are characterised
by $(\xbm \xcc),$ $(\xbm PR),$ and the additional property

$(\xbm CUM)$ $ \xbm (X) \xcc Y \xcc X$ $ \xcp $ $ \xbm (X)= \xbm (Y)$
 \xDH
Ranked relations are characterised by $(\xbm \xcc),$ $(\xbm PR),$
and the additional property

$(\xbm =)$ $X \xcc Y,$ $ \xbm (Y) \xcs X \xEd \xCQ $ $ \xcp $ $ \xbm (X)=
\xbm (Y) \xcs X.$
 \xEj

For more explanation and details, see e.g.
 \cite{Sch18}, in particular Table 1.6 there.

\efa

 \xEj

\vspace{3mm}



\vspace{3mm}

\clearpage



\begin{thebibliography}{xxxxxx}

\addcontentsline{toc}{section}{References}


\bibitem[Han69]{Han69}
B. Hansson, ``An analysis of some deontic logics'', Nous 3, 373--398.
Reprinted in R. Hilpinen, ed. ``Deontic Logic: Introductory and Systematic
Readings'', Reidel, pp. 121--147, Dordrecht 1971

\bibitem[KLM90]{KLM90}
S. Kraus, D. Lehmann, M. Magidor, ``Nonmonotonic reasoning, preferential
models and cumulative logics'', Artificial Intelligence, 44 (1--2),
pp. 167--207, July 1990.

\bibitem[Key21]{Key21}
J. M. Keynes, ``A treatise on probability'', London, 1921

\bibitem[LMS01]{LMS01}
D. Lehmann, M. Magidor, K. Schlechta, ``Distance semantics for belief
revision'', Journal of Symbolic Logic, Vol. 66, No. 1,
pp. 295--317, March 2001

\bibitem[Lew73]{Lew73}
D. Lewis, ``Counterfactuals'', Blackwell, Oxford, 1973

\bibitem[SEP13]{SEP13}
``Analogy and analogical reasoning'', Fall 13 edition,
Stanford Encyclopedia of Philosophy, 2013

\bibitem[SEP19c]{SEP19c}
``Analogy and analogical reasoning'',
Stanford Encyclopedia of Philosophy, 2019

\bibitem[Sch04]{Sch04}
K. Schlechta, ``Coherent systems'', Elsevier, Amsterdam, 2004.

\bibitem[Sch18]{Sch18}
K. Schlechta, ``Formal Methods for Nonmonotonic and Related
Logics'',
Vol. 1: ``Preference and Size'',
Vol. 2: ``Theory Revision, Inheritance, and Various Abstract
Properties''
Springer, 2018

\bibitem[Sta68]{Sta68}
R. Stalnaker, ``A theory of conditionals'', N. Rescher (ed.), ``Studies in
logical theory'', Blackwell, Oxford, pp. 98--112

\end{thebibliography}
\end{document}